\documentclass[twocolumn,amsmath,aps,fleqn, superscriptaddress]{revtex4}
 
\usepackage{amssymb}
\usepackage{amsmath}
\usepackage{epsfig}
\usepackage{subfigure}
\usepackage{mathrsfs}
\usepackage{longtable}
\usepackage{enumerate} 
\usepackage{multirow}
\usepackage{color}

\usepackage[usenames,dvipsnames]{xcolor}
\usepackage{hyperref}
\hypersetup{
    colorlinks = true,
    citecolor = {MidnightBlue},
    linkcolor = {BrickRed},
    urlcolor = {BrickRed}
}

\newcommand{\be}{\begin{eqnarray}}
\newcommand{\ee}{\end{eqnarray}}

\newcommand{\mnras}{MNRAS~}
\newcommand{\jcap}{JCAP~}
\newcommand{\apjl}{ApJL~}

 \allowdisplaybreaks[1]
 
\begin{document}

\title{{Spatial curvature endgame: \\ Reaching the limit of curvature determination}}

\author{C. Danielle Leonard}
\email{danielle.leonard@physics.ox.ac.uk}
\affiliation{Astrophysics, University of Oxford, Denys Wilkinson Building, Keble Road, Oxford, OX1 3RH, UK}

\author{Philip Bull}
\affiliation{California Institute of Technology, Pasadena, CA 91125, USA}
\affiliation{Jet Propulsion Laboratory, California Institute of Technology, 4800 Oak Grove Drive, Pasadena, California, USA}

\author{Rupert Allison}
\affiliation{Astrophysics, University of Oxford, Denys Wilkinson Building, Keble Road, Oxford, OX1 3RH, UK}

\date{\today}

\begin{abstract}
Current constraints on spatial curvature show that it is dynamically negligible: $|\Omega_{\rm K}| \lesssim 5 \times 10^{-3}$ (95\% CL). Neglecting it as a cosmological parameter would be premature however, as more stringent constraints on $\Omega_{\rm K}$ at around the $10^{-4}$ level would offer valuable tests of eternal inflation models and probe novel large-scale structure phenomena. This precision also represents the ``curvature floor'', beyond which constraints cannot be meaningfully improved due to the cosmic variance of horizon-scale perturbations. In this paper, we discuss what future experiments will need to do in order to measure spatial curvature to this maximum accuracy. Our conservative forecasts show that the curvature floor is unreachable -- by an order of magnitude -- even with Stage IV experiments, unless strong assumptions are made about dark energy evolution and the $\Lambda$CDM parameter values. We also discuss some of the novel problems that arise when attempting to constrain a global cosmological parameter like $\Omega_{\rm K}$ with such high precision. Measuring curvature down to this level would be an important validation of systematics characterisation in high-precision cosmological analyses.
\end{abstract}

\maketitle

\section{Introduction}
\label{sec:introduction}
\noindent
The question of whether spatial curvature is an important contribution to the cosmic energy budget has lately seemed all but settled. Current constraints from combined cosmic microwave background (CMB) and baryon acoustic oscillation (BAO) data find $|\Omega_{\rm K}| < 5 \times 10^{-3}$ (95\% CL) \cite{Planck2015}. The implication is that curvature is dynamically negligible, affecting cosmic expansion by less than 1\% at any epoch.

Is it time, then, to close the door on curvature, fixing it to zero in our cosmological analyses (as is already common practice)? In some contexts, this is certainly a valid choice -- for example, the effects of non-zero $\Omega_{\rm K}$ on the growth rate of structure are essentially negligible at the precision of today's experiments. However, to assume flatness exclusively would preclude a number of potentially powerful tests of early Universe physics, and of general relativistic effects in large-scale structure.

Constraints on $\Omega_{\rm K}$ at around the $10^{-4}$ level offer a stringent test of eternal inflation \cite{Kleban2012, 2012PhRvD..86b3534G, Guth2014}. Slow-roll eternal inflation predicts a strong bound on $|\Omega_{\rm K}|<10^{-4}$, while false-vacuum eternal inflation would be ruled out if $\Omega_{\rm K} < -10^{-4}$. Measuring a `large' $\Omega_{\rm K}$ would have profound implications for this important class of models. Inflationary scenarios that give rise to bubble collisions and other large-scale anomalies also tend to have observable levels of spatial curvature \citep{2015PhRvD..91l3523A, 2015arXiv150803786J}, and an open Universe ($\Omega_{\rm K} > 0$) has been proposed as a strong prediction of the string multiverse (although see \citep{2008PhLB..660..382B} for a refutation of this statement). There is therefore a clear theoretical motivation for seeking a curvature constraint at the 0.01\% level.

\begin{figure}[t]
\hspace{-0.5em}\includegraphics[width=\columnwidth]{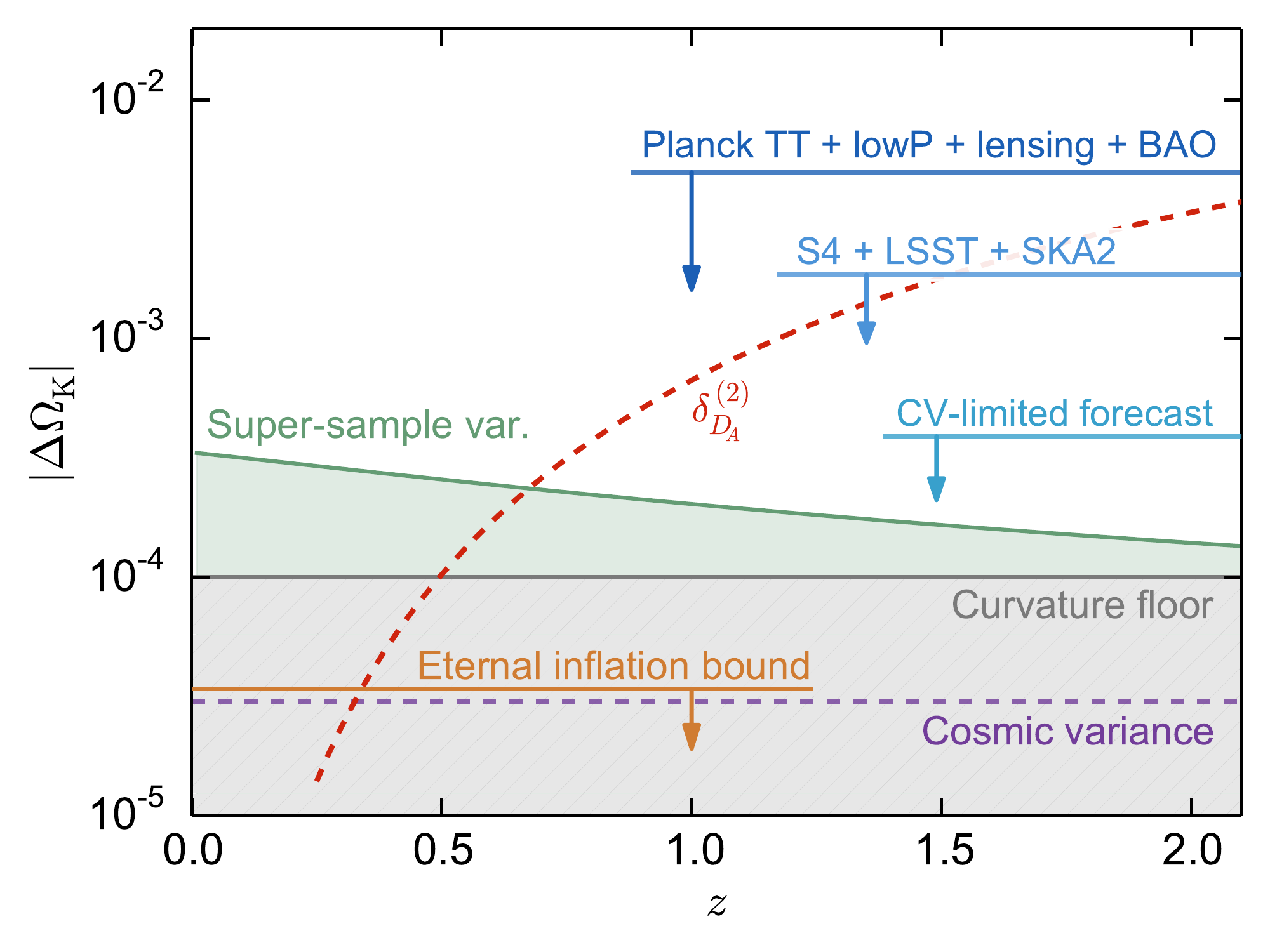}
\vspace{-1em}
\caption{Various contributions to the observed spatial curvature or its variance (shown as a function of redshift where relevant). Current and forecast upper limits on $\Omega_{\rm K}$ are shown as blue bars (note that the {\sl Planck} limit fixed $w$). The green shaded region shows the approximate super-sample variance $\sigma({\Omega_{\rm K}}) \sim \mathcal{O}(1) \times \sqrt{\langle \delta_b^2 \rangle}$ \citep{2014PhRvD..89h3519L} for a $(5\, {\rm Gpc})^3$ redshift bin, as a function of its centre redshift. The shift in $\Omega_{\rm K}$ (absolute value) due to the second-order correction to $D_A$ \citep{2015JCAP...06..050B} is shown as a red dashed line, and the maximum permissible value of $\Omega_{\rm K}$ in eternal inflation from \citep{Kleban2012} as an orange bar. The grey band is the curvature floor, and the purple dashed line is the cosmic variance limit from \citep{2008arXiv0804.1771W}. All upper limits and variance bounds are shown at the 95\% CL level.}
\label{fig:scales}
\vspace{-2em}
\end{figure}

Another motivation for this target is that $\sim\!\! 10^{-4}$ represents a ``floor'' below which $\Omega_{\rm K}$ cannot be decisively distinguished from primordial fluctuations.
The expected variance of curvature perturbations with wavelengths of order the horizon size represents an irreducible cosmic variance ``noise'' level, $\sigma(\Omega_{\rm K}) \approx 1.5 \times 10^{-5}$, below which increased observational precision cannot improve the constraint \cite{2008arXiv0804.1771W}. Framed as a Bayesian model selection problem, model confusion between curved and flat models actually becomes unavoidable at a higher threshold: $|\Omega_{\rm K}| \approx 10^{-4}$ if one demands ``strong'' evidence on the Jeffreys scale \cite{Vardanyan2009}. We adopt this more stringent value as the ``curvature floor''.

Additionally, an observation of non-zero spatial curvature could be the result of large-scale structure effects. Perturbations to the distance-redshift relation at second-order contribute a monopole at the sub-percent level, for example \citep{2015JCAP...06..050B}, leading to a shift in the apparent value of $\Omega_{\rm K}$. Local inhomogeneities contribute to the monopole too, with observers inside potential wells seeing shifts in $\Omega_{\rm K}$ \citep{Bull2013, 2015JCAP...07..025W, Valkenburgh2013}. In large-scale structure surveys, super-sample modes (those with wavelengths larger than the survey size) also contribute an apparent shift in the background cosmology \citep{2014PhRvD..90j3530L, 2014PhRvD..90b3003M}. More subtle general relativistic effects related to the behaviour of curvature in inhomogeneous spacetimes \citep{2016IJMPD..2530007B}, such as the non-commutation of spatial averaging with directional averaging \cite{2015JCAP...07..040B}, time evolution \cite{2010arXiv1003.4020V, 2012PhRvD..85d3506C}, and wide-angle effects \citep{2016arXiv160309073D}, can also lead to observable discrepancies between different measurements of curvature. Finally, certain alternative theories of gravity predict a non-zero observed $\Omega_{\rm K}$ \cite{Barrow2011}.

Given that current curvature upper limits are 1--2 orders of magnitude away from the level required to probe most of these effects, there is an imperative to continue pushing $\Omega_{\rm K}$ constraints to greater precision. In this paper we address the question of when, and how, we can expect to make cosmological observations that will detect or constrain $\Omega_{\rm K}$ at the $10^{-4}$ level. Recent efforts have explored future constraints within a range of observational scenarios and analysis frameworks \cite{2016arXiv160309073D, Knox2006, Knox2006c, Zhan2008, Mortonson2009, Vardanyan2009, Barenboim2010, Vardanyan2011, Smith2012, Sapone2014, Takada2015, Chen2016}. The focus in these previous forecasts has largely been on geometric observables, although several probes of the growth of structure have also been considered. We expand upon these efforts by considering combined constraints from the CMB, BAO, and the weak gravitational lensing of galaxies. The former two probes are arguably the `purest' precision observables, in that they are likely to offer the best control over systematic effects and biases. We have similarly selected weak gravitational lensing on the basis that it is a key observable of several upcoming surveys, and hence an intensive study of relevant systematic errors is currently underway.

There are two principle ways in which we seek to improve upon previous forecasts. First, we take a broadly conservative approach. We incorporate parameters that may exhibit significant degeneracies with $\Omega_{\rm K}$ (e.g. the neutrino mass and the time-evolution of dark energy), and then examine the effect of varying or fixing these parameters. We also fold in uncertainties due to observational nuisance parameters, and comment on several possible additional sources of systematic bias. In contrast, most previous work has focused upon single or limited extensions beyond a non-flat $\Lambda$CDM model. Second, we explore a suite of current and upcoming surveys in combination, whereas previous work has generally focused upon a single set or very limited sets of future surveys. In this way, we aim to answer the question of how and when we might first achieve the target constraint of $10^{-4}$, rather than to examine the properties of a particular survey or surveys of interest.

The paper is structured as follows. In Sec.~\ref{sec:setup}, we describe the observational probes and our forecasting methodology. In Sec.~\ref{sec:physics}, we present forecast spatial curvature constraints for three generations of ongoing and upcoming surveys, identifying the combination of surveys most likely to reach the curvature floor first, while highlighting sources of systematic bias that could jeopardise the measurement. We conclude in Sec.~\ref{sec:conclusion} with a discussion of the implications of our results for tests of inflation and of large-scale structure effects, as well as for the next generation of cosmological surveys.

\section{Surveys and forecasting method}
\label{sec:setup}
\subsection{Observational probes}
We examine the constraints that can be placed on spatial curvature by three observational probes: CMB, BAO, and weak gravitational lensing. We will use information from the power spectra of these observables only, neglecting three-point (and higher) correlations in the galaxy distribution, for example.

Looking first to measurements of the CMB, we consider the angular power spectra of the temperature and polarisation anisotropies. We also include information from the CMB lensing convergence power spectrum, the theoretical form of which is given below in Eq.~\ref{pkappa}.  

Measurements of the BAO scale are included using a formalism based on that presented in \cite{2007ApJ...665...14S}, in which the redshift-space galaxy power spectrum is modelled as
\begin{eqnarray}
P_g(k, \mu) = \left ( b + f \mu^2 \right )^2 P_{\rm sm}(k) \left [ 1 + f_{\rm BAO}(k) \right ] \nonumber\\
\times e^{-\frac{k^2}{2}([1 - \mu^2]\Sigma^2_\perp + \mu^2 \Sigma^2_\parallel)}, ~~~~
\label{PgBAO}
\end{eqnarray}
where $b$ and $f$ are the linear bias and linear growth rate factors \citep{1987MNRAS.227....1K}, $P_{\rm sm}$ is the smooth (BAO-free) power spectrum, and $f_{\rm BAO}$ contains the scale information of the BAO feature. To extract only the BAO information, we relabel the argument of $f_{\rm BAO}$ as $k^\prime$, such that we consider only derivatives with respect to the BAO scale in the Fisher analysis described below. $k^\prime$ is defined as
\be
(k^\prime)^2 = \alpha_s^2 \left [ (\alpha_\perp k_\perp)^2 + (\alpha_\parallel k_\parallel)^2 \right ],
\ee
where $k_\perp = k \sqrt{1 - \mu^2}$, $k_\parallel = k\mu$, $\{\alpha_\parallel, \alpha_\perp\}$ represent shifts with respect to a fiducial cosmology, and $\alpha_s$ accounts for uncertainty in modelling shifts of the BAO scale due to non-linearities. We marginalise over $b$ and $f$, but assume that the BAO smoothing is known, so $(\Sigma_\perp, \Sigma_\parallel)$ are treated as fixed parameters.

Our third observable is the weak gravitational lensing of galaxy images by large-scale structure. The angular power spectra of the convergence for galaxy lensing, CMB lensing, and their cross-correlation, are given by the form
\begin{align}
C^{A^i B^j}_\ell&=\frac{9}{16}\int_{0}^{\chi_\infty}d\chi \frac{g^{A^i}(\chi)g^{B^j}(\chi)}{f_K(\chi)^2c^3} (H(\chi)a(\chi))^4  \nonumber \\ &~~~~~~~~~~~~~~~~~~~~~\times \Omega_{\rm M}(\chi)^2 P_{\delta}\left(\frac{\ell}{f_K(\chi)}, \chi \right),
\label{pkappa}
\end{align}
where $A$ and $B$ indicate galaxies or CMB, while $i$ and $j$ refer to the source galaxy redshift bin (relevant only in the galaxy case). Note that we employ the Limber approximation \cite{Limber1953, Simon2007}, and so consider a minimum multipole of $\ell_{\rm{min}}=10$ for all lensing spectra in this work. 

In Eq.~\ref{pkappa}, $g^{A^i}(\chi)$ is the lensing kernel,
\begin{equation}
g^{A^i}(\chi) = 2 f_K(\chi) \int_{\chi}^{\chi_\infty} \frac{f_K(\chi'-\chi)}{f_K(\chi')}W^{A^i}(\chi') d\chi'.
\end{equation}
In the case of CMB lensing, $W^{\kappa_c}=\delta(\chi - \chi_*)$, where $\chi_*$ is the comoving distance to last scattering. For galaxy lensing, $W^{\kappa_g^i}(\chi)$ is the distribution of source galaxies in each redshift bin $i$, and accounts for photometric redshifts using the method described in \cite{JoachimiSchneider2009}. The photo-z bias is assumed to be zero, while the photo-z scatter, $\sigma_z$, is used to determine the number of redshift bins such that $\Delta z = 3\sigma_z(1+z)$. Values of $\sigma_z$ are given for each weak lensing survey in Table \ref{table:WLsurveys}. The true redshift distribution of the total population of source galaxies is modelled using the form $n(z) \propto z^{\alpha} \exp[-({z/z_0})^\beta]$ from \cite{Smail1994}, where $z_0=z_m / 1.412$, and $z_m$ is the median redshift. 

We also account for the possibility of intrinsic galaxy alignments, which contaminate the observed galaxy ellipticity. For the galaxy lensing auto-correlation, we have
\begin{align}
C^{\epsilon^i \epsilon^j}_\ell &= C^{\kappa_g^i \kappa_g^j}_\ell + C^{\kappa_g^i I^j}_\ell + C^{I^i \kappa_g^j}_\ell + C^{I^i I^j}_\ell
\label{IAform}
\end{align}
where $I$ represents the intrinsic ellipticity. For the cross-correlation between galaxy and CMB lensing, there is a similar adjustment, which we compute as described in \cite{2014MNRAS.443L.119H}: $C^{\epsilon^i \kappa_c}_\ell = C^{\kappa_g^i \kappa_c}_\ell + C^{I^i \kappa_c}_\ell$.

We base our expressions for the final three terms of Eq.~\ref{IAform} on those from \cite{Kirk2011}. However, where they assume that all galaxies contribute equally to the intrinsic alignment signal, we follow \cite{Chisari2015} and assume that only red galaxies contribute. We additionally make the simplifying assumption that the fraction of red galaxies $f_{\rm{red}}$ is constant over the redshifts that we consider. The result is that each of the final three terms of Eq.~\ref{IAform} depends on an amplitude parameter $f_c$, where $f_c = C_1  \rho_c f_{\rm{red}}$, and $C_1$ is the standard amplitude parameter for intrinsic alignments. We marginalise over the combined parameter $f_c$ in our forecasts to account for uncertainty in the intrinsic alignment amplitude.  

We also expect some minor cross-correlation between CMB temperature and CMB lensing via the ISW effect. This is comparatively negligible, however.

\begin{center}
\begin{table}
\begin{tabular}{| c | c | c | c | c | c | c | c |}
\hline
& $\alpha$ & $\beta$ & $z_0$ & $\sigma_z$ & $n_{\rm gal}$ & $ \langle \gamma_{\rm int}^2 \rangle^{\frac{1}{2}}$ & $ f_{\rm sky}$ \\
\hline
\,\,DES\,\, &\,\,\,2\,\,\,&\,\,\,1.5\,\,\,&\,\,\,0.425\,\,\,&\,\,\,0.07\,\,\,&\,\,\,12\,\,\,&\,\,\,0.32\,\,\,&\,\,\,0.12\,\,\,\\
\,\, Euclid\, \, &\,\,\,2\,\,\,&\,\,\,1.5\,\,\, &\,\,\,0.637\,\,\,&\,\,\,0.05\,\,\,&\,\,\,30\,\,\,&\,\,\,0.22\,\,\,&\,\,\,0.375\,\,\,\\
\,\,LSST\,\,&\,\,\,2\,\,\,&\,\,\,1\,\,\,&\,\,\,0.5\,\,\,&\,\,\,0.03\,\,\,&\,\,\,40\,\,\,&\,\,\,0.18\,\,\,&\,\,\,0.485\,\,\,\\
\hline
\end{tabular}
\vspace{-0.2em}
\caption{Survey parameters for the three generations of weak gravitational lensing surveys considered. $n_{\rm gal}$ is given in units of galaxies per square arcminute.}
\vspace{-1.2em}
\label{table:WLsurveys}
\end{table}
\end{center}
\subsection{Fisher forecasting methodology}
\label{subsec:fisher}

We explore the ability of current and future experiments to improve constraints on $\Omega_{\rm K}$ by using a Fisher forecasting methodology (see for example \cite{Fisher1935,Bassett2011}). The inverse Fisher information matrix approximates the covariance matrix for an experiment, given a fiducial signal model and its behaviour as a function of selected free parameters, as well as the experiment's noise characteristics. The level of optimism in Fisher forecasting can be controlled by accounting for various nuisance parameters which would be introduced in a realistic analysis.

We compute the Fisher matrix for each experiment with respect to the parameters: $\Omega_{\rm K}$, $\Omega_{\rm B} h^2$, $\Omega_{\rm C} h^2$, $h$, $n_s$, $A_s$, $\tau$, $M_\nu$, $w_0$, $w_a$, $\{f_i\}$, $\{b_i\}$, and $f_c$. As can be seen from this parameter list, a non-$\Lambda$CDM expansion history is permitted. Fiducial values of $\Lambda$CDM parameters are taken as reported by {\sl Planck} in 2015 \cite{Planck2015}, while $\{w_0, w_a\}$ are taken as $-1$ and $0$, respectively. The fiducial linear bias in each redshift bin, $b_i$, is survey-specific, and $f_i$ is the linear growth rate per bin. $f_c$ is fiducially taken as $0.0067$, following the standard convention in which $C_1 \rho_c = 0.0134$ \cite{Chisari2015} and setting $f_{\rm{red}}$ fiducially to $0.5$. 

Note that Fisher matrices containing independent information can be directly summed in order to obtain a combined Fisher matrix. Here, we assume three independent Fisher matrices: one for CMB temperature and polarisation, one for BAO, and one for CMB lensing and galaxy lensing. By computing the Fisher matrices in this manner, any covariance between CMB temperature/E-mode polarisation and lensing has been neglected.

The single Fisher matrix describing both the temperature and polarisation of the CMB takes the form
\begin{equation}
\mathcal{F}_{ab}^{\rm CMB} = \sum_{\ell} f_{\rm sky} \left(\partial_a C_\ell\right)^T \Sigma_\ell^{-1} (\partial_b C_\ell), 
\label{CMBfisher}
\end{equation}
where $f_{\rm sky}$ is the fractional sky coverage, and $\Sigma_\ell$ is the covariance matrix between angular power spectra at a given $\ell$. $\Sigma_\ell$ has dimensions $N\times N$ where $N$ is equal to the number of spectra considered. $\partial_a C_\ell$ is a vector of length $N$, containing derivatives of each spectrum at multipole $\ell$ with respect to parameter $a$. 

Similarly, the weak gravitational lensing of galaxies and of the CMB are described in a single Fisher matrix. We employ the formalism developed in \cite{Hu1999}:
\begin{equation}
\mathcal{F}_{ab}^{\rm L}=\sum_{\ell}
\frac{2\ell+1}{2}f_{\rm sky}\mathrm{Tr}\left[(C_\ell)^{-1}\partial_aC_\ell \,(C_\ell)^{-1}\partial_bC_\ell\right].
\label{Fishertomo}
\end{equation}

$C_\ell$ is a square matrix with dimensions of the number of galaxy redshift bins plus one (for the CMB). Each element of the $C_\ell$ matrix provides the sum of the theoretical auto- or cross-spectrum and the relevant noise.
\begin{center}
\begin{table*}[t]
{
\begin{tabular}{| l || r | r || r | r | r | r | r | r || r || r |}
\hline
{\bf Experiments} & \begin{tabular}[c]{@{}c@{}}Uninform. \\priors\end{tabular} & Mild priors & Fixed $w$ & Fixed $\alpha_s$ & Fixed $M_\nu$ & Fixed $\tau$ & Fixed $f_c$ & Fixed $b_i$ & \begin{tabular}[c]{@{}c@{}}$\ell_{\rm max}=$ \\2000\end{tabular}&Fixed all \\
\hline
{\sl Planck} CMB & 393.4 & 280.7 & 239.3 & 280.7 & 258.4 & 267.9 & 280.7 & 280.7 & 280.7 & 4.6\\
~~+ BOSS BAO & 382.0 & 144.2 & 59.5 & 144.1 & 138.1 & 142.7 & 144.2 & 144.2 & 144.2 & 4.6\\
~~+ DES WL & 312.9 & 240.4 & 228.6 & 240.4 & 219.6 & 232.8 & 220.4 & 240.4 & 188.8 &4.6 \\
~~+ both & 305.9 & 118.0 & 56.4 & 117.8 & 114.1 & 116.7 & 118.0 & 118.0 & 105.8 & 4.6\\
\hline
Advanced ACTPol CMB & 164.1 & 128.7 & 116.1 & 128.7 & 76.3 & 123.7 & 128.7 & 128.7 & 128.7 & 1.2\\
~~+ Euclid BAO & 153.6 & 44.1 & 18.8 & 43.4 & 29.1 & 41.0 & 44.1 & 44.1 & 44.1 & 1.2\\
~~+ Euclid WL & 97.9 & 83.4& 70.2 & 83.4 & 43.0 & 80.7 & 74.4 & 83.4 & 42.7 & 1.2\\
~~+ both & 87.7 & 25.3 & 18.3 & 23.2 & 23.3  & 23.5 & 24.0 & 25.3 & 19.7 & 1.2\\
\hline
S4 CMB & 94.1 & 74.9 & 63.6 & 74.9 & 39.2 & 73.4 & 74.9 & 74.9 & 74.9 & 0.9\\
~~+ SKA2 BAO & 68.2 & 31.4 & 13.9 & 30.6 & 21.7 & 28.5 & 31.4 & 31.4 & 31.4 & 0.9 \\
~~+ LSST WL & 56.6 & 51.8 & 31.2 & 51.8 & 23.6 & 51.2& 45.0& 51.8 & 24.7 & 0.9 \\
~~+ both & 47.8 & 22.2 & 12.7 & 19.4 & 17.3 & 21.3 & 21.1 & 22.2 & 15.0 & 0.9 \\
\hline
CV-limited & 3.6& 3.5 & 3.3 & 2.2 & 3.5 & 1.9 & 3.4 & 3.5& ---~ & 0.4\\
\hline
\end{tabular} }
\caption{Forecast marginal constraints on $\Omega_{\rm K}$ (95\% CL), divided by $10^{-4}$, for all experiment combinations, and for various priors on other parameters. In columns (3-9), the mild priors of the second column are assumed for the non-fixed parameters (see text). The final column shows the results when no other parameters are marginalised over. Note that there is no result for the CV-limited combination of surveys in the $\ell_{\rm max}=2000$ column because we assume $\ell_{\rm max}=5000$ for weak lensing in this case.}
\label{table:results}
\end{table*}
\end{center}

Finally, the Fisher matrix of BAO is related to that for galaxy clustering. In the distant observer approximation,
\begin{eqnarray}
\mathcal{F}_{ab} ^{\rm Gal}&=& \int_{-1}^{+1} d\mu \int_{k_{\rm min}}^{k_{\rm max}} \frac{dk}{(2\pi)^2} k^2 V_{\rm sur} \left ( \frac{P_g(k, \mu)}{P_g(k, \mu) + 1/n} \right )^2 \nonumber\\
&& ~~~~~~~~~~~~~~~~~~~~~~~\times (\partial_a \log P_g) (\partial_b \log P_g),
\end{eqnarray}

\noindent where $n$ is the number density of detected galaxies. In order to extract information only from the BAO, derivatives are considered only with respect to the BAO scale $k^\prime$ described above, i.e.: $\partial_\alpha P_g \propto (\partial_{k^\prime} f_{\rm BAO}(k^\prime)) \partial_\alpha k^\prime$. The desired Fisher matrix, $\mathcal{F}_{ab}^{\rm BAO}$, then follows directly.

\subsection{Numerical issues and non-linear scales}
\label{subsec:num}
We use the public CAMB code \cite{Lewis:1999bs} to output CMB temperature and polarisation spectra, as well as the matter power spectra necessary to compute BAO and lensing observables using the expressions above. Given our observables, we then compute all derivatives numerically. 

When employing numerical derivatives, it is crucial to select a step size in the differentiation parameter which lies within the regime of convergence -- too small a step can yield numerical errors, while too large will depart from the regime of validity. In the case of galaxy weak lensing, ensuring convergence proved non-trivial. This was due to the prescription for computing the non-linear matter power spectrum: in the case where non-linearities were included, derivative convergence was not uniformly achievable at higher multipoles, but when predictions were artificially restricted to linear theory, convergence was easily achieved. Therefore, in order to be certain of robust results, we report constraints from galaxy weak lensing using $\ell_{\rm max}=300$. We select this maximum multipole because it provides agreement of better than $5\%$ between the problematic non-linear case and the well-behaved linear case. Note that we make this conservative $\ell_{\rm max}$ cut only for galaxy lensing and for the cross-correlation between galaxy and CMB lensing; the CMB lensing auto-correlation is less affected due to its sensitivity to higher redshifts, where non-linear effects are less important. 

Although we make this cut to deal with numerical problems, we note that it also serves our overall goal of providing conservative forecasts, as it naturally excises multipoles at which non-linearities become important. Regardless of numerical issues, at smaller scales we would be faced with uncertainties in non-linear modelling and baryonic physics that currently affect the matter power spectrum at around the $10\%$ level. To illustrate the potential of including higher multipoles were this problem to be solved in the future, we also present constraints with $\ell_{\rm max}=2000$ (and, in the case of the hypothetical cosmic-variance-limited lensing survey explored below, with $\ell_{\rm max}=5000$). However, these constraints may be subject to errors due to the numerical issues discussed above, so should be treated with care.

\subsection{Cosmological surveys}
\label{subsec:surveys}
\noindent
We compute Fisher matrices for experiments that are representative of three generations of cosmological surveys, for CMB, weak lensing, and BAO observations:
\begin{itemize}
\item{{\bf Stage II (current):} {\sl Planck} \cite{Planck2015}, the Dark Energy Survey (DES) \cite{DES}, and the Baryon Oscillation Spectroscopic Survey (BOSS) \cite{Dawson2012}.}
\item{{\bf Stage III (next generation):} Advanced ACTPol (Atacama Cosmology Telescope) \cite{Henderson2015}, and Euclid \cite{Refregier2010} (for both galaxy lensing and BAO).}
\item{{\bf Stage IV (future):} A Stage IV CMB survey \cite{Wu2014}, the Large Synoptic Survey Telescope (LSST) \cite{LSST}, and Stage 2 of the Square Kilometre Array \cite{Maartens:2015mra}.}
\end{itemize}
In defining these three generations of experiments, we use nomenclature similar to that of the Dark Energy Task Force \cite{Albrecht2006}, but do not attempt to match their stages exactly. For example, Euclid is technically a Stage IV experiment, but we include it in Stage III due to its earlier operating timeframe than SKA2. Similarly, DES is included in Stage II despite typically being considered a Stage III experiment. SKA2 is selected as the Stage IV BAO experiment due to the fact that it is expected to outperform non-radio counterparts in this observable out to $z=1.4$ \cite{Bull2015}. We select Advanced ACTPol as the Stage III CMB experiment, but mention SPT-3G (South Pole Telescope) \cite{Benson2014} as another possible choice.

The survey parameters that we used in our Fisher forecasting are given in Table \ref{table:WLsurveys} for weak lensing surveys, while those pertaining to BAO were given in \cite{2014JCAP...05..023F, 2013LRR....16....6A, 2016ApJ...817...26B} for BOSS, Euclid, and SKA2 respectively. The specifications employed here for Advanced ACTPol are the same as those given in \cite{2015PhRvD..92l3535A} for the Stage III (wide) experiment, while the CMB Stage IV survey considered here employs the specifications of the Stage IV CMB survey described in the same work.  Note that, in all cases, forecast constraints from CMB experiments include CMB lensing as well as polarisation.

\section{Results}
\label{sec:physics}
\noindent

For each `generation' of survey, we consider four combinations of observables: CMB-only, CMB + BAO, CMB + galaxy weak lensing (WL), and the combination of all three. The results are shown in Table~\ref{table:results}, which presents the forecast 95\% CL constraint on $\Omega_{\rm K}$ for each combination, and for a range of different prior assumptions. Note that values in the table have been divided by  $10^{-4}$.

\subsection{Forecasts for each generation of surveys}
\label{sec:forecasts}

We begin with the most pessimistic case, where all parameters are marginalised without reference to any prior information (the ``uninformative priors'' column in Table~\ref{table:results}). Our {\sl Planck} + BOSS forecast predicts a $95\%$ CL constraint on $\Omega_{\rm K}$ of approximately $3.8\times10^{-2}$, nearly eight times worse than the published {\sl Planck} + BAO constraint \cite{Planck2015}. This is the result of a geometric degeneracy \citep{2007JCAP...02..001I, 2007JCAP...08..011C, 2007PhRvD..76j3533W, Farooq2015}, which makes it difficult to disentangle the effects of curvature and a varying dark energy equation of state. Information from CMB lensing and low-redshift BAO helps to break the degeneracy, but the parameters remain strongly correlated; if $w_0$ and $w_a$ are fixed (see the ``fixed $w$'' column), the constraint is very similar to the published result (which also assumed fixed $w$). Examining the ``uninformative priors'' column more generally, we see significant improvement in the constraint on $\Omega_{\rm{K}}$ with progressive generations. The most powerful $95 \%$ CL constraint, coming from the combination of all three Stage IV experiments, is $4.8 \times 10^{-3}$, far from the target precision of $\sim 10^{-4}$.

A more realistic (but still conservative) analysis corresponds to the ``mild priors'' column in Table~\ref{table:results}. In this case, we assumed that external Gaussian priors would be applied to certain parameters, corresponding to 95\% (2$\sigma$) bounds of $\sigma(\alpha_s) = 0.01$, $\sigma(b_i) = 1$, $\sigma(M_\nu) = 0.4$, $\sigma(f_c) = 0.05$, $\sigma(w_0) = 1$, $\sigma(w_a) = 2$, and $\sigma(\tau) = 0.1$. These priors are chosen to be mostly uninformative (i.e. weak), except for helping to break the more severe degeneracies. External probes (e.g. supernovae in the case of $w_0$ and $w_a$, or 21cm experiments in the case of $\tau$ \citep{2016PhRvD..93d3013L}) are capable of providing external constraints at this level, without suffering from the same degeneracies. These mild priors offer varying levels of improvement over the case of uninformative priors. The most dramatic gain is for the Advanced ACTPol + Euclid BAO combination of surveys, where the constraint is tightened by approximately a factor of 3.5.

Next, in each subsequent column, a particular parameter is fixed to its fiducial value in an attempt to understand its individual effect on the spatial curvature constraint. As discussed above, fixing the equation of state of dark energy (the ``fixed $w$'' column) can result in large improvements by removing a geometric degeneracy. However, a key goal of some of the experiments under consideration (e.g. Euclid) is to measure the time-dependence of $w$. For this reason, it is arguable that $w_0$ and $w_a$ should remain free. Still, if one had a strong theoretical prior on a cosmological constant, fixing these parameters would result in considerable gains for curvature constraints -- the S4 + SKA2 + LSST combination yields a 95\% bound of $1.3\times 10^{-3}$, about a factor of two better than in the ``mild priors'' case.

Fixing the neutrino mass, $M_\nu$, is shown to be only mildly helpful in increasing the combined constraining power of all three observables, although its effect is non-negligible for constraints from the CMB alone. Fixing $\tau$ has a similarly small effect, with improvements reflecting the breaking of the degeneracy between the optical depth and $A_{s}$, which is itself degenerate with $\Omega_{\rm K}$. We see, however, that fixing $\tau$ has a considerable effect in the case of the combination of cosmic-variance-limited surveys (discussed below), demonstrating the potential importance of this degeneracy for even more futuristic observations than those of Stage IV.

Fixing the BAO scale exactly (see the ``fixed $\alpha_s$'' column) has only a very mild effect for the Stage III and IV experiments. Similarly, fixing either the galaxy bias $b_i$ or intrinsic alignments parameter $f_{c}$ has a minimal effect on the forecast constraint values. These parameters are of more relevance as sources of potential bias in inferred parameters values, as will be discussed in Section \ref{sec:systematics}. 

Allowing the galaxy weak lensing power spectra to contribute out to a multipole of $2000$ rather than $300$ (`$\ell_{\rm max} = 2000$') results in a more noticeable effect: for Stage IV,  the improvement over the ``mild priors'' scenario in the three-observable case exceeds $\sim\!30\%$, reaching a constraint on $\Omega_{\rm K}$ of $1.5 \times 10^{-3}$.

The last column of Table~\ref{table:results} shows the constraints that would be achieved if $\Omega_{\rm K}$ was the only unknown parameter. In this case, $10^{-4}$ is achievable by S4 alone -- the CMB offers the best (conditional) constraint on $\Omega_{\rm K}$. Beyond fixing those parameters which have already been discussed, the greatest effect in producing these tight forecast constraints comes from fixing $h$, with a secondary but relevant effect from fixing $\Omega_{\rm C}$ and $\Omega_{\rm B}$. Controlling all other parameters to such a high precision is unrealistic, however -- the conclusion from our analysis is therefore that $10^{-3}$ (95\% CL) is the most likely achievable constraint on $\Omega_{\rm K}$ for the foreseeable future.

Finally, to explore what might be possible in the more distant future, we include also forecast constraints for the case where all three surveys are cosmic variance-limited. Interestingly, we see that even for this combination of highly-idealised surveys, only the case where all auxiliary parameters are fixed can provide a sub-$10^{-4}$ constraint on the spatial curvature. 

\subsection{Systematics, degeneracies, and theoretical uncertainties}
\label{sec:systematics}

We now briefly discuss some of the key modelling and parameter uncertainties that are likely to affect a precision curvature measurement.

\paragraph{Shifts in the BAO scale:} Non-linear evolution of the galaxy distribution broadens the BAO feature slightly, shifting the peak by $\sim 0.1 - 0.3\%$ \citep{2007ApJ...664..660E, 2010ApJ...720.1650S}. This shift can be partially undone by calibrating off simulations, and correcting for coherent peculiar velocities using the `reconstruction' technique \citep{2007ApJ...664..675E}\footnote{The statistical nature of the reconstruction procedure means that it does not always improve the precision of the distance measurement \citep{2014MNRAS.441.3524K}.}.
Uncertainty in the redshift evolution and scale-dependence of the galaxy bias also have a small effect on the peak position \citep{2011ApJ...734...94M}. One set of simulations to estimate the effects of non-linear evolution on the BAO peak measured the distribution of shifts to be $(0.3 \pm 0.015)\%$ at $z \simeq 0$ \citep{2010ApJ...720.1650S}. An observer could correct for the shift by marginalising over this distribution, which would correspond to setting a prior of $\sigma({\alpha_s}) \approx 0.015\%$ in our forecasts. Conservatively, we chose a looser $1\%$ prior for most columns in Table~\ref{table:results}, although it can be seen that this does not significantly change the results (and even fixing $\alpha_s$ has little effect).

\paragraph{Non-linear power spectrum:} Weak lensing is also sensitive to the modelling of the non-linear power spectrum. This is often dealt with by reducing $\ell_{\rm max}$, so that only more linear modes are used, at the cost of reducing the constraining power of a given experiment. As discussed above, this is the approach we take for most columns in Table~\ref{table:results}, choosing $\ell_{\rm max}=300$. The lensing power spectrum is more complicated to model than the BAO feature, so there is little one can do to improve this situation other than trying to model the non-linear matter power spectrum as accurately as possible. This requires high-precision simulations that include realistic baryonic effects \citep{2005APh....23..369H, 2011MNRAS.415.3649V, 2012JCAP...04..034H}; however, current uncertainties in baryonic modelling on non-linear scales are relatively large ($\sim 10\%$) \citep{2015JCAP...12..049S}. As demonstrated in column 9 of Table~\ref{table:results}, if more accurate modelling of the non-linear power spectrum were to allow us to increase $\ell_{\rm max}$ for galaxy lensing to $2000$, this could improve constraints on $\Omega_{\rm K}$ by $30\%$ for Stage IV surveys. (See \citep{2014arXiv1410.6826M} for a more optimistic assessment of the importance of baryonic effects in weak lensing observations.)

\paragraph{Massive neutrinos:} Spatial curvature and the sum of neutrino masses are correlated in CMB observations principally due to their similar effect on the amplitude of the CMB lensing power spectrum (see Figure 9 of \cite{2015PhRvD..92l3535A}). Increasing the neutrino mass increases the matter density and therefore enhances the growth of structure, which induces a larger lensing effect on the CMB. This can be compensated by increasing the curvature parameter $\Omega_{\rm K}$ which, for a fixed CMB acoustic angular scale, decreases the matter density and the corresponding lensing amplitude. It is therefore vital to allow for variations in the neutrino mass in any cosmological analysis involving spatial curvature (and vice versa).

\paragraph{Dark energy evolution:} As discussed above, there is a strong degeneracy between the dark energy equation of state and $\Omega_{\rm K}$; some of the redshift scaling of the curvature term can always be absorbed into a sufficiently unconstrained $w(z)$. A similar `dark degeneracy' also exists for the matter density, $\Omega_{\rm M}$ \citep{2009PhRvD..80l3001K}. This problem is often solved in an ad hoc way, by fixing $w$ (as in, for example, \citep{Planck2015}), but this is a strong choice of prior. A more conservative alternative may be to use theoretical priors on dark energy models, e.g. \citep{2014PhRvD..90j5023M, 2015JCAP...11..029P, 2016arXiv160208283S}. These employ stability conditions and physical modelling assumptions to establish a subset of ({\it a priori}) viable models from a broad class of dark energy theories. These theoretical priors can be surprisingly restrictive; using them, instead of allowing completely arbitrary functional forms for $w(z)$, one can hope to partially break the degeneracy with curvature in a more physically-justified manner.

\paragraph{Intrinsic alignments:} The intrinsic alignment of galaxies contributes to the observed galaxy lensing power spectrum, as well as to the observed cross-spectrum between galaxy lensing and CMB lensing. The choice of prior for the amplitude of the intrinsic alignment contribution has a small effect on the spatial curvature constraint, as described above. We do, however, find that increasing the fiducial value of the amplitude parameter $f_c$ leads to a tighter constraint on $\Omega_{\rm K}$. For example, for the combination of all three Stage III experiments, $\sigma(\Omega_{\rm K})$ decreases by $\sim\!6\%$ when the fiducial value for $f_{\rm red}$ is increased from $0.1$ to $1$ (with fixed fiducial $C_1$). This result is somewhat surprising, as we might expect that adding more `contaminant' to the lensing signal in the form of intrinsic alignments would loosen cosmological constraints. Intrinsic alignments are sensitive to cosmological parameters in their own right though, and so it is entirely plausible that an increase in their amplitude would render the lensing signal more sensitive to spatial curvature -- essentially, the IA contribution contains extra information about $\Omega_{\rm K}$.

\paragraph{Super-sample modes and local environment:} Density fluctuations on scales larger than the survey size couple to small-scale modes, causing shifts in observable quantities that are degenerate with a change in background cosmological parameters like $\Omega_{\rm K}$. Super-sample modes are a significant source of sample variance in weak lensing surveys, and can potentially cause a large degradation in parameter constraints \citep[e.g.][]{2009ApJ...701..945S, 2014IAUS..306...78T}. Their effects can again be calibrated using simulations \citep{2014PhRvD..89h3519L, 2015arXiv151101465B}, and the parameter degeneracies can be broken through measurements of the power spectrum covariance \citep{2014PhRvD..90j3530L}.

Inhomogeneities local to both the source and observer can also shift observables away from their background values, as well as contributing to the sample variance \citep{2006PhRvD..73b3523B}. A coherent local inhomogeneity, such as the potential well of the local supercluster \cite{2014Natur.513...71T}, can bias the inferred distance-redshift relation, again leading to a shift in the observed $\Omega_{\rm K}$. This can be corrected through sufficiently precise modelling of local structures, or high-precision CMB spectral distortion measurements \citep{Bull2013}.

\paragraph{Higher-order perturbations:} At the precision level being considered in this paper, higher-order corrections to perturbative quantities are not necessarily negligible. For example, at $z=1$, second-order lensing effects contribute a $\sim 8 \times 10^{-4}$ correction to the angular diameter distance \citep{2015JCAP...06..050B}, leading to a shift in $\Omega_{\rm K}$ significantly larger than the target uncertainty if left uncorrected. The form of the higher-order perturbations depends on the observable in question, but can be calculated exactly at a given order for a given set of background cosmological parameters \cite{2012JCAP...11..045B, 2013JCAP...11..019F, 2014CQGra..31t2001U}. A number of novel physical effects also arise at higher order and are worthy of further study in their own right \cite{2014CQGra..31t2001U}; high-precision curvature observations will necessarily measure some of them.

\paragraph{CMB systematics:} Systematic effects in space-based CMB experiments like {\sl Planck} are mostly well understood, at least in temperature maps \cite{2015arXiv150702704P}. However, there remain some systematics affecting the CMB lensing reconstruction and polarisation data that are not fully understood \cite{PlanckLensing2015}. Additionally, the data analysis challenges for forthcoming ground-based experiments are uncertain. Coherent fluctuations of the atmosphere may prove difficult to model, and could affect the sensitivity and CMB lensing estimation performance of experiments like Advanced ACT, especially on large scales. Polarised foregrounds are also proving harder to clean than initially expected \cite{2015PhRvL.114j1301B}. The most likely impact on $\Omega_{\rm K}$ constraints is to increase the errorbars by degrading their sensitivity.

\paragraph{Photometric redshifts:} We have assumed that the photometric redshift scatter $\sigma_z$ is perfectly known for each survey, and have fixed the photometric redshift bias to zero. In principle, an error could be introduced into the weak lensing spatial curvature constraints by uncertainty in either of these parameters. Work is ongoing to adequately calibrate photometric redshift measurements for current and future surveys (see, for example, \cite{Masters2015, Cunha2014}).

\section{Discussion and Conclusions}
\label{sec:conclusion}

We have shown that forthcoming surveys -- even the combination of Stage IV CMB + BAO + weak lensing experiments -- are likely to place constraints on the spatial curvature of $\sim 10^{-3}$ (95\% CL) at best. This is an order of magnitude worse than the `ultimate' precision on $\Omega_{\rm K}$ required to put constraints on eternal inflation and to detect several large-scale structure effects which induce an apparent spatial curvature.

This would at first glance seem to be at odds with some predictions in the literature, which have reported that constraints at the $\sim{\rm few} \times 10^{-4}$ level may be achievable even with single experiments, or when combined with {\sl Planck} CMB measurements (see, for example, \cite{Vardanyan2009, Takada2015}). Our approach has differed in that we have performed consistent and conservative forecasts for a selection of real (current or planned) surveys for three observables simultaneously -- BAO, CMB, and weak lensing -- each of which is expected to have precise control over systematic effects once the observations have fully matured. This is important, as even small systematic shifts in the observations could cause a spurious detection of curvature at the low level being probed here.

We have also incorporated a set of cosmological and nuisance parameters that cannot be neglected. As shown in the final column of Table~\ref{table:results}, the $\sim 10^{-4}$ level is reachable by both Stage III and Stage IV CMB experiments if all other parameters are held fixed. This situation is unrealistic, however. Even then, we have neglected to fold several other effects into our forecasts, such as super-sample variance and corrections from higher-order perturbation theory (see Sec.~\ref{sec:systematics}), which can be expected to contribute additional uncertainty in $\Omega_{\rm K}$.

This does not mean that the ``curvature floor'' is unreachable in principle. Other observational probes could improve on the constraints we have presented here, either directly, by measuring distances and the expansion rate more precisely, or indirectly, by helping to break parameter degeneracies. Experiments like Euclid and SKA2 may produce tighter measurements of $\Omega_{\rm K}$ by using information from redshift-space distortions and the broadband shape of the galaxy power spectrum, while Type Ia supernova samples will greatly increase in size in the coming years. Experiments targeting the epoch of reionization (e.g. through 21cm intensity mapping) will help to break the $\tau$ degeneracy \citep{2016PhRvD..93d3013L}, while radio weak lensing studies will improve our understanding of intrinsic alignments \citep{2016arXiv160103947H}. The systematic effects and modelling uncertainties affecting these probes are, however, typically worse, or currently less well-understood, than for the three used here, which may lead to concerns about the robustness of any $0.01\%$ constraint which depends on them.

This, really, is the big question in modern observational cosmology: how well we can hope to understand the myriad systematic and theoretical uncertainties that affect various cosmological observables, as well as low-level corrections (such second-order effects) that are simply unobservable in current data. In other words, how {\it accurate} can our cosmological inferences be, given their impressive forthcoming precision?

Spatial curvature, with its relatively well-understood physical causes and clear target precision level, represents an `acid test' for this level of accuracy in cosmology. Reaching the curvature floor, and agreeing on the interpretation of whatever we see there, will be a definitive sign of maturity for the field -- whenever we get there.

\begin{acknowledgements}
We would like to thank Pedro Ferreira and Jo Dunkley for helpful discussions. We also thank the authors of CAMB, which was used in this work. CDL is supported by the Natural Sciences and Engineering Research Council of Canada. PB's research was supported by an appointment to the NASA Postdoctoral Program at the Jet Propulsion Laboratory, California Institute of Technology, administered by Universities Space Research Association under contract with NASA. RA is supported by ERC grant 259505.
\end{acknowledgements}



\end{document}